# Anomalous gravitational vacuum fluctuations which act as virtual oscillating dipoles

Giovanni Modanese
*University of Bolzano*
*Italy*
*Inst. for Advanced Research in the Space,*
*Propulsion & Energy Sciences*
*Madison, AL, U.S.A.*

## 1. Introduction

In this work we would like to review some concepts developed over the last few years: that the gravitational vacuum has, even at scales much larger than the Planck length, a peculiar structure, with anomalously strong and long-lasting fluctuations called "zero-modes"; and that these vacuum fluctuations behave as virtual particles of negative mass and can interact with each other, leading to the formation of weakly bound states. The bound states make up a continuum, allowing at each point of spacetime the local excitation of the gravitational vacuum through the coupling with matter in a coherent state. The spontaneous or stimulated decay of the excited states leads to the emission of virtual gravitons with spin 1 and large $p/E$ ratio. The main results on the zero-modes and their properties have been given in (Modanese, 2011), but in this work we expand and discuss in physical terms several important details concerning the zero-mode interactions, the dynamics of virtual particles with negative mass and the properties of virtual gravitons.

Technically, our approach is based on the Lorenzian path integral of Einstein gravity in the usual metric formulation. We take the view that any fundamental theory of gravity has the Einstein action as its effective low-energy limit (Burgess, 2004). An important feature of the path integral approach is that it allows a clear visualization of the metric as a dynamical quantum variable, of which one can study averages and fluctuations also at the non-perturbative level. It is hard, however, to go much further than formal manipulations in the Lorenzian path integral; after proving the existence of the zero-modes we resort to semi-classical limits and standard perturbation theory. This method is clearly not always straightforward. At several points we proceed, by necessity, through physical induction and analogies with other interactions.

The outline of the work is the following. In Section 2 we show the existence of the zero-modes and discuss their main features, using their classical equation and the path integral. This Section contains some definitely mathematical parts, but we have made an effort to translate all the concepts in physical terms along the way. Section 3 is about the pair interactions of zero-modes: symmetric and antisymmetric states, transitions between these states, virtual dipole emission and its *A* and *B* coefficients. Section 3.3 contains a digression on the elementary dynamics of virtual particles with negative mass. Section 4 is devoted to the interaction of the zero-modes with a time-variable Λ-term. The effect of this term is compared to that of "regular" incoherent matter by evaluating their respective transition rates. Finally, in Section 5 we discuss in a simplified way the properties of virtual gravitons; the virtual gravitons exchanged in a quasi-static interaction are compared to virtual particles exchanged in a scattering process and to virtual gravitons emitted in the decay of an excited zero-mode.

## 2. Isolated zero-modes: non trivial static metrics with null action

Our starting point is a very general property of Einstein gravity: it has a non-positive-definite action density. As a consequence, some non trivial static field configurations (metrics) exist, which have zero action. We call these configurations zero-modes of the action. The Einstein action is $S_E = -\frac{c^4}{8\pi G}\int d^4x\sqrt{g}R$ (plus boundary term; see Sect. 3) and the zero-mode condition is

$$\int d^4x\sqrt{g}R = 0 \tag{1}$$

This condition is, of course, satisfied by any metric with $R(x)=0$ everywhere (vacuum solutions of the Einstein equations (28), like for instance gravitational waves). But since the density $\sqrt{g}R$ is not positive-definite, the condition can also be satisfied by metrics which do not have $R(x)=0$ everywhere, but regions of positive and negative scalar curvature.

We are interested into these configurations because, in the Feynman path integral, field configurations with the same action tend to interfere constructively and so to give a contribution to the integral distinct from the usual classical contribution of the configurations near the stationary point of the action. Let us write the Feynman path integral on the metrics $g_{\mu\nu}(x)$ as

$$I = \int d[g]\exp\left(\frac{i}{\hbar}S_E[g]\right) \tag{2}$$

Suppose there is a subspace $X$ of metrics with constant action. The contribution to the integral from this subspace is simply

$$I_X = \exp\left(\frac{i}{\hbar}\hat{S}_E\right)\int_X d[g] = \exp\left(\frac{i}{\hbar}\hat{S}_E\right)\mu(X) \tag{3}$$

where $\hat{S}_E$ is the constant value of the action in the subspace and $\mu(X)$ its measure. The case $\hat{S}_E = 0$ is a special case of this.

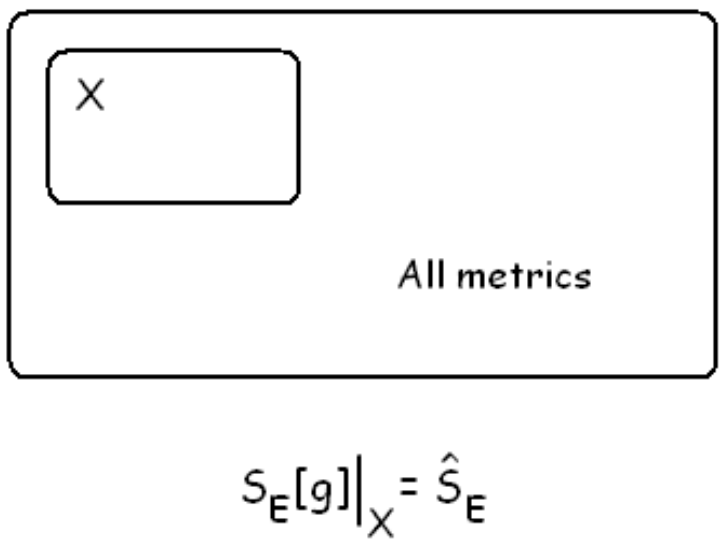

Fig. 1. Subspace *X* of metrics with constant action. All the metrics (spacetime configurations) in *X* have the same action $\hat{S}_E$. In particular, there exist a subspace whose metrics all have zero action.

The zero-modes can only give a significant contribution to the path integral if they are not isolated configurations (like a line in 2D, which has measure zero), but a whole full-dimensional subset of all the possible configurations. They are "classical" fields, not in the sense of being solutions of the Einstein equations in vacuum, but in the sense of being functions of spacetime coordinates which are weighed in the functional integral with non-vanishing measure.

### 2.1 Classical equation of the zero-modes

Now let us find at least some of these configurations. It is not obvious that eq. (1) has solutions with *R* not identically zero, because it is a difficult non-linear integro-differential equation.

In some previous work we used, to solve (1) in the weak field approximation, a method known as "virtual source method" or "reverse solution of the Einstein equations" (Modanese, 2007). According to this method, one solves the Einstein equations with non-physical sources which satisfy some suitable condition, in our case $\int dx\sqrt{g}g^{\mu\nu}T_{\mu\nu} = 0$. Since for solutions of the Einstein equations one has (trace of the equations) $R = \frac{8\pi G}{c^4}g^{\mu\nu}T_{\mu\nu}$, it follows that such solutions will be zero-modes. The expression $\int dx\sqrt{g}g^{\mu\nu}T_{\mu\nu} = 0$ is far simpler in the linear approximation. In that case the source must satisfy a condition like, for instance, $\int dxT_{00} = 0$ (supposing $T_{ii}$ is vanishing) and is therefore a "dipolar" virtual source.

A much more interesting class of zero-modes is obtained, however, in strong field regime, starting with a spherically-symmetric Ansatz. In other words, let us look for spherically symmetric solutions of (1). Consider the most general static spherically symmetric metric

$$d\tau^2 = B(r)dt^2 - A(r)dr^2 - r^2(d\theta^2 + \sin^2\theta d\phi^2) \tag{4}$$

where $A(r)$ and $B(r)$ are arbitrary smooth functions. We add the requirement that outside a certain radius $r_{\text{ext}}$, $A(r)$ and $B(r)$ take the Schwarzschild form, namely

$$B(r) = \left(1 - \frac{2GM}{c^2 r}\right); \quad A(r) = \left(1 - \frac{2GM}{c^2 r}\right)^{-1} \quad \text{for } r \geq r_{ext} \tag{5}$$

This requirement serves two purposes: (1) It allows to give a physical meaning to these configurations, seen from the outside, as mass-energy fluctuations of strength $M$. For $r>r_{\text{ext}}$ their scalar curvature is zero. (2) More technically, the Gibbons-Hawking-York boundary term of the action is known to be constant in this case (Modanese, 2007).

Even with only the functions $A$ and $B$ to adjust, the condition (1) is very difficult to satisfy. We do find a set of solutions, however, if we make the drastic simplification $g_{00}=B(r)=const$. The scalar curvature multiplied by the volume element becomes in this case

$$L = \sqrt{g}R = -8\pi\sqrt{|BA|}\left(\frac{rA'}{A^2} + 1 - \frac{1}{A}\right) \tag{6}$$

Apart from the constant $c^4/8\pi G$, $L$ is the lagrangian density of the Einstein action, computed for this particular metric. Let us fix arbitrarily a reference radius $r_{\text{ext}}$, and introduce reduced coordinates $s=r/r_{\text{ext}}$. Define an auxiliary function $\alpha=A^{-1}$. Regarding $L(s)$ as known, eq. (6) becomes an explicit first-order differential equation for $\alpha$:

$$\alpha' = \frac{1}{s} - \frac{\alpha}{s} + \frac{L\sqrt{|\alpha|}}{8\pi s\sqrt{|B|}} \tag{7}$$

The boundary conditions (5) are written, in reduced coordinates

$$B(s \geq 1) = \left(1 - \frac{\tilde{M}}{s}\right); \quad A(s \geq 1) = \left(1 - \frac{\tilde{M}}{s}\right)^{-1} \tag{8}$$

where $\tilde{M}$ is a free parameter, the total mass in reduced units: $\tilde{M} = 2GM / c^2 r_{ext}$. In the following we shall take $\tilde{M} < 0$, in order to avoid singularities. For $r < r_{ext}$, we have $B = B(1) = 1 - \tilde{M}$.

It is interesting to note that putting $L$=0 in eq. (7) we can easily find an exact solution, ie a non-trivial static metric with $R$=0. Namely, if 1-α>0, then $\alpha=1-e^{\text{const}}/s$, which does not satisfy the boundary condition; if 1-α<0, then $\alpha=1+e^{\text{const}}/s$, implying $e^{cost}=-\tilde{M}$ . The resulting $g_{rr}$ component has the same form on the left and on the right of $s$=1, namely

$$g_{rr}=\left(1+\frac{|\tilde{M}|}{s}\right)^{-1} \tag{9}$$

while $g_{00}$ is constant and equal to $(1+|\tilde{M}|)$ for $s$<1, and is equal to $(1+|\tilde{M}|/s)$ for $s\geq 1$. Note that $g_{rr}$ goes to zero at the origin.

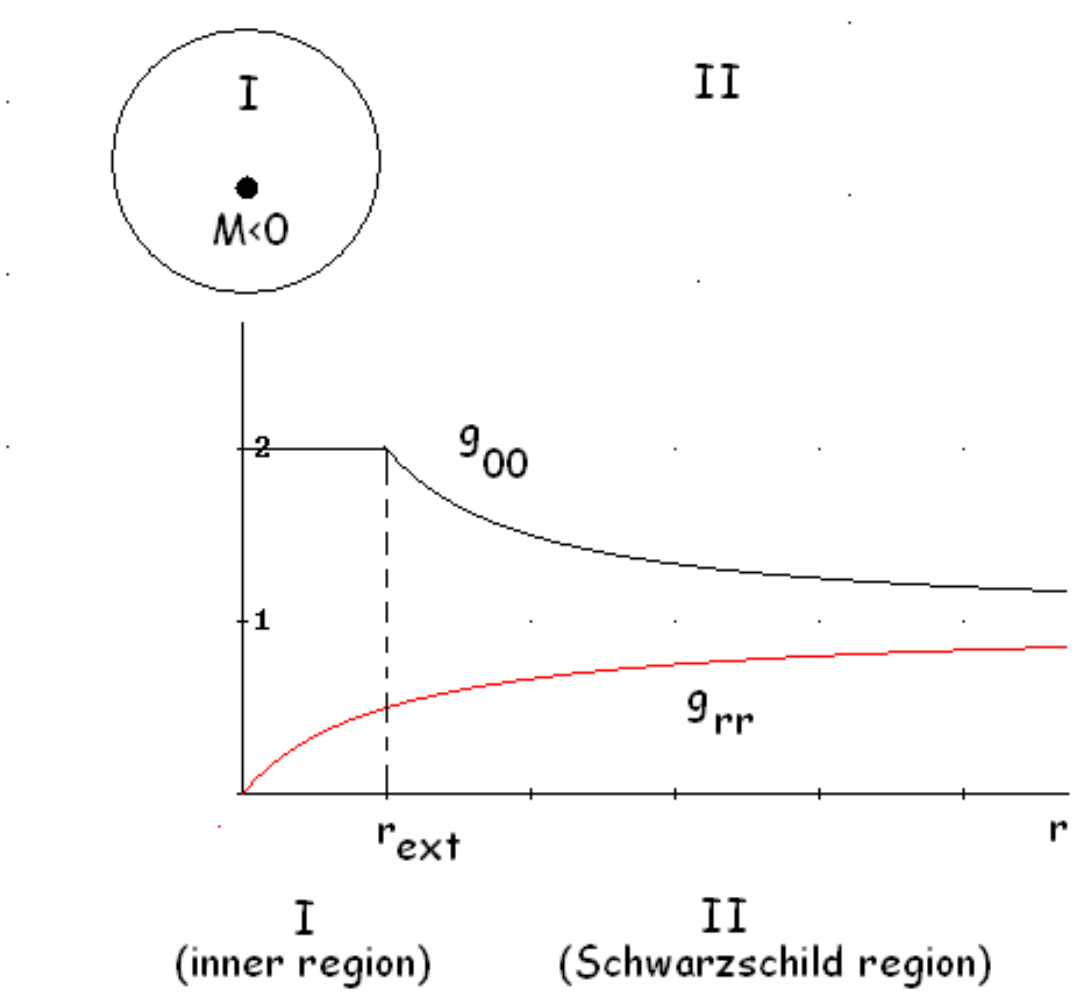

Fig. 2. Metric of an elementary static zero-mode of the Einstein action. Inside the radius $r_{ext}$ (region I) the $g_{00}$ component is constant, and the $g_{rr}$ component goes to zero. On the outside (region II) both components have the form of a Schwarzschild solution with negative mass.

Now we can look for metrics close to (9), but with scalar curvature not identically zero. For large $\tilde{M}$ and small $L$, the last term in eq. (7) is a small perturbation. Since α never diverges and $\alpha^{-1}$ does not appear in the equation, the perturbed solution is not very different from (7). For values of $\tilde{M}$ of order 1 or smaller, the equation can be integrated numerically. If we choose a function $L(s)$ with null integral on the interval (0,1), we obtain a metric which is a zero-mode of the action but not of the lagrangian density. One can take, for instance, $L(s)=L_0\sin(2\pi ns)$, with $n$ integer.

In conclusion, we have found a family of regular metrics with null scalar curvature, depending on a continuous parameter $\tilde{M}$ . Furthermore, we have built a set of metrics close to the latter, by solving eq. (7) with $L$ arbitrary but having null integral. These metrics do not have zero scalar curvature, but still have null action. They make up a full-dimensional subset of the functional space (see proof in (Modanese, 2007)).

Our solutions of the zero-mode condition are, outside the radius $r_{ext}$, Schwarzschild metrics with $M$<0. The quantity $Mc^2$ coincides with the ADM energy of the metrics. At the origin of the coordinates the component $g_{rr}$ goes to zero, the integral of $\sqrt{g}R$ is finite and also the volume $\int dx\sqrt{g}$ is finite. The volume inside the radius $r_{ext}$ is smaller than the volume of a sphere with the same radius in flat space.

According to our previous argument on the functional integral, these metrics give a significant contribution to the quantum averages, although they are neither classical solutions nor quantum fluctuations near the classical solutions. In the vacuum state, there exists a finite probability that the metric at any given point is not flat, but has the form of a zero-mode, i.e., seen from a distance, of a pseudo-particle of negative mass. In the language of Quantum Field Theory this could be called a vacuum fluctuation. Vacuum fluctuations are created spontaneously and at zero energetic cost at any point of spacetime, in a homogeneous and isotropic way. Usually vacuum fluctuations have a very short life, as can be shown through the Schroedinger equation (time-energy uncertainty principle) or through a transformation to Euclidean time (when the action is positive-definite). These arguments on the lifetime of the fluctuations can not be applied here, because quantum gravity has neither a local Hamiltonian, nor a positive-definite action. Our fluctuations, if they were completely isolated, would be independent of time; in fact, their interaction causes a finite lifetime (Sect. 2.3). In Sect. 5 we shall give a comparison between this kind of vacuum fluctuations and other fluctuations present in quantum gravity, like the virtual gravitons which transmit the gravitational interactions.

In order to avoid a large global curvature, the *total* average effect of the virtual masses of the zero-modes must inevitably be renormalized to zero. This is, in our view, guaranteed by the "cosmological constant paradox": nature appears to be endowed with a dynamical mechanism which relaxes to zero any constant positive or negative contributions to the vacuum energy density, coming from particle physics or even from gravity itself. So, even though such contributions are formally infinite, in the end they do not affect the curvature of spacetime. The full explanation of this mechanism can only be achieved within a complete non-perturbative theory of Quantum Gravity. Some partial evidence of the dynamical emergence of flat spacetime has been obtained in the lattice theory, and in effective field theory approaches (Hamber, Dolgov …).

Therefore we shall not be concerned with the global effect of our massive vacuum fluctuations on spacetime. We shall instead consider their interactions, which result in a novel pattern of purely gravitational excited states, above a ground state in which all fluctuations pairs with equal mass are in a symmetrical superposition. Freely speaking, it's a bit like studying the local effects of pressure variations, without worrying about how the total force due to atmospheric pressure affects the Earth.

### 2.2 Zero-modes in the explicit functional integral

The zero-modes equation (plus the argument of non-interference) tell us that relevant run-away configurations of vacuum exist, in which the metric is locally very different from its classical value $g_{\mu\nu}(x)=\eta_{\mu\nu}$.We shall now consider an explicit path integral of Einstein gravitation, in order to evaluate the functional average of certain metric components and confirm this supposition.

Let us choose a spherical coordinate system. We integrate only over a sector $X$ of the functional space, namely over the spherically-symmetric metric configurations with constant $g_{00}$. If we obtain a null quadratic vacuum average in $X$, namely

$$\langle g_{rr}(0)\rangle_X=\frac{\int_X d[g]\exp\left\{\frac{i}{\hbar}S[g]\right\}g_{rr}(0)}{\int_X d[g]\exp\left\{\frac{i}{\hbar}S[g]\right\}}=0 \tag{10}$$

this allows us to reach our conclusion: at any point there is a finite probability for a zero-mode to occur.

For these metrics the Einstein action is written (Sect. 2.1)

$$S_E=-\frac{c^4}{8\pi G}\int d^4x\sqrt{g(x)}R(x)=\frac{4\pi c^4}{G}\int dt\int_0^\infty dr\sqrt{|BA|}\left(\frac{rA'}{A^2}+1-\frac{1}{A}\right) \tag{11}$$

where $A=g_{rr}$ and $B=g_{00}$ are functions of $r$. Define a radius $r_{ext}$, the "external radius" of our configurations, on which we impose boundary conditions as in Sect. 2.1. This means that we integrate over configurations which outside the radius $r_{ext}$ appear like Schwarzschild metrics with mass $M$. In order to avoid singularities, we suppose $M<0$. We can re-write the action as an integral on $r$ with upper limit $r_{ext}$, because the scalar curvature of the Schwarzschild metric is zero. We can also add the Gibbs-Hawking-York boundary term, which in this case takes the form $S_{HGY}=-M\int dt$. For a fixed time interval, we can regard the integral $\int dt$ as a constant.

Supposing $B$ constant ($B=1-\tilde{M}$), the path integral over these field modes is written

$$\begin{aligned}&\int d[A]\exp\left\{\frac{i}{\hbar}(S_E+S_{HGY})\right\}=\\&=\int d[A]\exp\left\{\frac{i}{\hbar}\frac{4\pi\sqrt{|B|}}{G}\int dt\int_0^1 ds\sqrt{|A|}\left(\frac{sA'}{A^2}+1-\frac{1}{A}\right)\right\}\exp\left\{-\frac{i}{\hbar}M\int dt\right\}\end{aligned} \tag{12}$$

The second exponential can be disregarded in the functional averages, because it cancels with the normalization factor in the denominator. In the first exponential, let us define a constant factor $\alpha = \frac{1}{\hbar} \frac{4\pi\sqrt{|B|}}{G} \int dt$ and discretize the integral in *ds*. We divide the integration interval [0,1] in $(N+1)$ small intervals of length $\delta$ and replace the integral with a sum, where the derivative is written as a finite variation. We obtain

$$\int d[A] e^{\frac{i}{\hbar} S_E} = \int \prod_{j=0}^{N+1} dA_i \exp\left\{ i\alpha\delta \sum_{j=0}^{N} \sqrt{|A_j|} \left( \frac{j\delta\left(A_{j+1} - A_j\right)}{\delta A_j^2} + 1 - \frac{1}{A_j} \right) \right\} \tag{13}$$

The presence of the square root and of the fractions with $A_j$ makes the integrals very complicated. Let us change variables. Suppose $A > 0$, which is physically a widely justified assumption (and remember we are looking for a sufficient condition, i.e. we want to show that there exist a set of gravitational configurations for which the functional average of a quadratic quantity is different from the classical value). Define $\gamma = 1/\sqrt{A}$. This gives the new path integral

$$\int_0^{+\infty} \prod_{j=0}^{N+1} d\gamma_j \frac{2}{\gamma_j^3} \exp\left\{ -i\alpha\delta \sum_{j=0}^{N} \left( 2j\delta \frac{\left(\gamma_{j+1} - \gamma_j\right)}{\delta} - \frac{1}{\gamma_j} + \gamma_j \right) \right\} \tag{14}$$

(Note that $\left(\gamma_j + \gamma_{j+1}\right) \approx 2\gamma_j$ in the continuum limit.) We want to use this to compute the average $\left\langle \gamma_m^2 \right\rangle$, where *m* is a fixed intermediate index. This is the average of the squared field $\gamma^2$ at the point $s = m\delta$, therefore in the continuum limit it gives the average of $\gamma^2$ at the origin. We know that the system has zero-modes for which $A \to 0$ at the origin, and therefore $\gamma \to \infty$. So we would like to show that $\left\langle \gamma_m^2 \right\rangle \to \infty$ for $\delta \to 0$. This can indeed be done (Modanese, 2011), and implies in turn that (10) is true. One can also check that this is not an artefact of the continuum limit.

### 2.3 Zero-modes as quantum states

The explicit calculation of the average $\left\langle g_{rr}(0) \right\rangle_X$ in a sector of the functional integral is conceptually important, but in practice it does not help much in giving a quantum representation of the zero-modes and their interactions. The properties of the zero-modes as "classical" metrics are more useful for that purpose. We shall suppose that each zero-mode corresponds to a quantum state $|i\rangle$ and that $\left\langle i \middle| H \middle| i \right\rangle = c^2 M_i$ (see below for the meaning of the gravitational Hamiltonian *H* in this context). The states $|i\rangle$ are localized and mutually orthogonal. Different $|i\rangle$ correspond to field configurations centered at different points. In the following we shall also suppose for simplicity that their Schwarzschild radii are always much smaller than their distance.

According to this line of thought, the "true non-interacting ground state" of the gravitational vacuum is obtained in principle as the limit of an infinite incoherent superposition of flat spacetime (Fock vacuum) plus single zero-mode wavefunctions:

$$|0\rangle = |0\rangle_{Fock} + \sum_i \xi_i \, |i\rangle \tag{15}$$

This definition of the ground state is clearly difficult to put on a rigorous basis. We are mainly interested, however, into the *excitations* with respect to this ground state. The most relevant among these excitations are those resulting from pair interactions of zero-modes, as we shall see.

Note that fixing $\langle i|H|i\rangle$ amounts to a much weaker statement than giving a gravitational quantum Hamiltonian operator *H*, because $\langle i|H|i\rangle$ is only a matrix element and a classical limit of the total energy for an asymptotically flat configuration (ADM energy). So whenever we write here the full gravitational Hamiltonian *H*, in fact we only exploit some properties of its matrix elements, like in a Heisenberg representation of quantum mechanics. This is consistent with our path integral approach to the full-interacting case.

In other words, in the following we use neither the "full" gravitational Hamiltonian operator *H*, nor eigenvalue relations. (Interaction Hamiltonians on a background metric like that employed in Sect. 4 do not suffer from these limitations.) In fact, the Hamiltonian *H* is very difficult to define in quantum gravity. Even classically, there exists no generally accepted expression for the gravitational energy density. Furthermore, assuming the validity of eigenvalue operator relations would lead to contradictions. For instance, by applying the full Hamiltonian to the vacuum state (15) and supposing for a moment that $H|i\rangle = M_i c^2 |i\rangle$, we would obtain, *only formally*

$$\text{"} H|0\rangle = \sum_i \xi_i M_i c^2 |i\rangle \text{"} \tag{16}$$

From this we would conclude that $|0\rangle$ is not an eigenstate! Nevertheless the property $\langle 0|H|0\rangle = 0$ is true, considering that the coefficients $\xi_i$ have random phases.

We could call the states $|i\rangle$ "purely gravitational, long-lived virtual particles". They are long-lived in the following sense. The classical equation for isolated zero-modes gives configurations independent from time. Adding to the pure Einstein action the boundary Gibbs-Hawking-York term, the latter takes the form $S_{GHY} = -M\int dt$, i.e. it is a constant for any fixed time interval, and does not cause interference in the path integral. However, when the zero-modes are not isolated but interact with each other, the boundary term causes their lifetime to be finite.

In the next section we shall discuss the simplest interaction of the zero-modes (pair interaction). This displays one of the typical amazing features of virtual particles (compare Sect. 5): they are created from the vacuum "for free", but after that they follow the usual dynamical rules. When computing the amplitude of a process involving virtual particles, we do not need to take into account the initial amplitude for creating the particles at a given point of space and time, but we do compute (Sect.s 3 and 4) the amplitudes for their ensuing propagation and interaction.

## 3. Pair interactions of zero-modes

We have introduced the concept of ground state in an effective theory of Quantum Gravity as given by the Fock vacuum plus a random superposition of zero-modes. In this Section we show that non-interacting zero-modes with equal mass are coupled in degenerate symmetric and anti-symmetric wavefunctions. The introduction of interaction removes the degeneration. The excited states form a continuum and the interaction of the vacuum with an external coherent oscillating source leads to transitions, with a probability which we shall compute in Sect. 4. As in Sect. 2, we denote with a capital $M$ a zero-mode mass (virtual and negative).

### 3.1 Pairs in symmetric and antisymmetric states

Consider a couple of states $|1\rangle$ and $|2\rangle$ with masses $M_1$ and $M_2$. We have

$$\langle 1|H|1\rangle = c^2 M_1 |1\rangle, \quad \langle 2|H|2\rangle = c^2 M_2 |2\rangle, \quad \langle 1|2\rangle = 0 \tag{17}$$

Putting now $M_1=M_2=M$ and taking the interaction into account, the degenerate non-interacting levels are splitted. Define the symmetrical and anti-symmetrical superpositions $\psi^+$ and $\psi^-$:

$$|\psi^+\rangle = \frac{1}{\sqrt{2}}(|1\rangle + |2\rangle) \quad |\psi^-\rangle = \frac{1}{\sqrt{2}}(|1\rangle - |2\rangle) \tag{18}$$

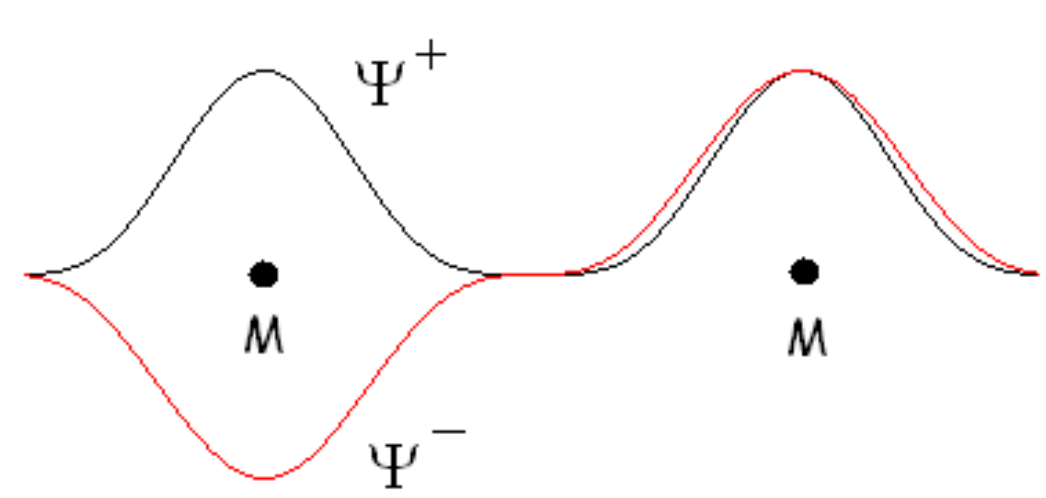

Fig. 3. Symmetric and antisymmetric bound states of zero-modes with equal mass *M*. (We assume that the wavefunction is much more localized near the masses than depicted – compared to their distance.)

The energy splitting $\Delta E$ is given, as known, by

$$\Delta E = E_- - E_+ = \langle \psi^- | H | \psi^- \rangle - \langle \psi^+ | H | \psi^+ \rangle = -2\langle 1 | H | 2 \rangle \tag{19}$$

Note that the matrix element $\langle 1 | H | 2 \rangle$ can be taken to be real without loss of generality. Suppose that $\langle 1 | H | 2 \rangle$ can be computed to a first approximation through its classical limit. The ADM energy integral at spatial infinity for the Schwarzschild-like field of two positive masses can be analytically continued to negative masses. We then obtain

$$\Delta E = 2\frac{GM^2}{r} \tag{20}$$

being *r* the distance between the symmetry centers of the states $|1\rangle$ and $|2\rangle$. This procedure reminds the computation of the bound states of two atoms in a molecule: the "internal states" of the atoms are not relevant and each atom is described by a single vector coordinate; the relevant Hamiltonian is the interaction Hamiltonian, although the full Hamiltonian of the system comprises in principle the forces inside the atoms and even inside the nuclei.

Let us consider the transitions between $\psi^+$ and $\psi^-$. We shall see that they are mainly of two types: (a) excitation $\psi^+ \to \psi^-$ due to the interaction with a local $\Lambda$-term dependent on time (variable vacuum energy density, associated with coherent matter - compare Sect. 4); (b) decay $\psi^- \to \psi^+$ with emission of a virtual graviton. We look for a relation between the frequency of the transition and the virtual mass of the excited states. In the ground state, all couples with equal mass will be in their symmetric superposition state. Any transition of one couple from its symmetric to its antisymmetric state gives an excited state with energy (20). Since there exist zero-modes with any (negative) mass, at any distance, there is actually a continuum of excited states.

For the same energy, in principle, there are transitions to excited levels involving different masses at different distances, provided the ratio $M^2/r$ is the same. In practice, however, there is an upper limit on the scale *r*, because the time-variable Λ-term has a typical spatial extension (coherence range) of the order of $10^{-9}$ m, and typical frequency $10^6$-$10^9$ Hz. This fixes the maximum virtual mass involved, by eq. (20), to $M \approx 10^{-13}$ kg. This is small, but definitely much larger than any atomic scale mass, and implies that also the gravitational interaction in the pairs of virtual masses is much larger than the usual gravitational interactions at atomic scale.

We are confronted here with a very unusual situation and we should check that our description is consistent, at least at the energy scale we are considering. (In principle the zero-mode fluctuations exist at any scale, but since they are an emergent phenomenon, computed in an effective theory, it is fair to concentrate on the scale which we deem most realistic.) First, one can easily check that the supposed localization of the zero-modes is well compatible with the Heisenberg position-momentum uncertainty principle. Second, one can prove that their interaction, though strong on the atomic scale, is much weaker than the interaction in a hypothetical gravitational bound state formed by two masses of this size. This can be easily checked, for instance, by computing the corresponding Bohr radius: this is of the order of $\hbar^2/Gm^3 \approx 10^{-19}$ m, while the zero-modes in the states $\psi^+$ and $\psi^-$ are separated by a distance of the order of $10^{-9}$ m. So the acceleration of each zero-mode due to the presence of the other is very small, if compared to accelerations due to atomic or molecular forces. It follows that in these "weakly bound states of heavy quasi-particles" the distance *r* varies slowly and there is plenty of time for the transitions $\psi^+ \leftrightarrow \psi^-$ to occur at frequency $10^6$-$10^9$ Hz, as we shall describe in detail later.

On a longer time scale, the interaction itself causes the zero-modes to fade out slowly as vacuum fluctuations. This is a subtle point that completes our analysis of the isolated zero-modes given in Sect. 2. As we have seen, the boundary term $M\int dt$ in the action is constant for an isolated zero-mode, for any time interval, and therefore an isolated zero-mode will persist indefinitely in time. For interacting zero-modes the situation is more complicated, because

(1) The superposition of their metrics is not necessarily a zero-mode.

(2) Their total ADM mass-energy is still constant, as long as radiation is negligible; this total mass-energy comprises their masses plus potential and kinetic energy. But when the emitted radiation becomes a sizeable fraction of the total mass, the ensuing change in the boundary term in the action of the zero-modes begins to cause a destructive interference in the functional integral between the metrics $g_{\mu\nu}(x,t_1)$, $g_{\mu\nu}(x,t_2)\ldots$ at subsequent times. So the quantum amplitudes of these metrics tend to vanish and the result is that the zero-modes, as vacuum fluctuations, acquire a finite lifetime as they begin to emit dipolar or quadrupolar radiation.

### 3.2 Virtual dipole emission, *A* and *B* coefficients

In this Section we compute the lifetime of an excited state $\psi^-$. The decay of the excited state occurs with the emission of an off-shell graviton with spin 1. This happens because the dominant graviton emission process in the decay of an excited zero-mode is oscillating-dipole emission. Quadrupolar emission, which is the only process ensuring conservation of energy, momentum and spin in the emission of on-shell gravitons, can in this case be disregarded. Since we are only interested into a lowest-order perturbative estimate (tree diagrams) we can use the linearized Einstein theory in the form of the "Maxwell-Einstein" equations

$$\begin{aligned}
&\nabla\cdot\mathbf{E}_G = -4\pi G\rho_m \\
&\nabla\times\mathbf{E}_G = -\frac{\partial\mathbf{B}_G}{\partial t} \\
&\nabla\cdot\mathbf{B}_G = 0 \\
&\nabla\times\mathbf{B}_G = -\frac{4\pi G}{c^2}\mathbf{j}_m + \frac{1}{c^2}\frac{\partial\mathbf{E}_G}{\partial t}
\end{aligned} \tag{21}$$

Here $\mathbf{E}_G$ is the gravito-electric (Newtonian) field, $\mathbf{B}_G$ is the gravito-magnetic field, and $\mathbf{j}_m$, $\rho_m$ are the mass-energy current and density. The elementary quantization of the field modes in a finite volume *V* follows the familiar scheme used for the computation of spontaneous and stimulated electromagnetic emission of atoms in a cavity. We have discussed in (Modanese, 2011) the conditions for applicability of the Einstein-Maxwell equations to plane waves in vacuum.

The Einstein *A*-coefficient of spontaneous emission turns out to be related to the *B*-coefficient and to the mass dipole moment by the relation

$$A = \left( B\frac{8\pi\hbar}{\lambda^3} \right) = \left( \frac{G}{\hbar^2} |\langle\hat{\mathbf{d}}\rangle|^2 \frac{8\pi\hbar}{\lambda^3} \right) \tag{22}$$

where the electromagnetic coupling constants have been replaced, up to an irrelevant adimensional factor, by the gravitational constants, according to eq.s (21). The operator $\hat{\mathbf{d}}$ is the mass-dipole moment and the matrix element is taken between the initial and final state of interest.

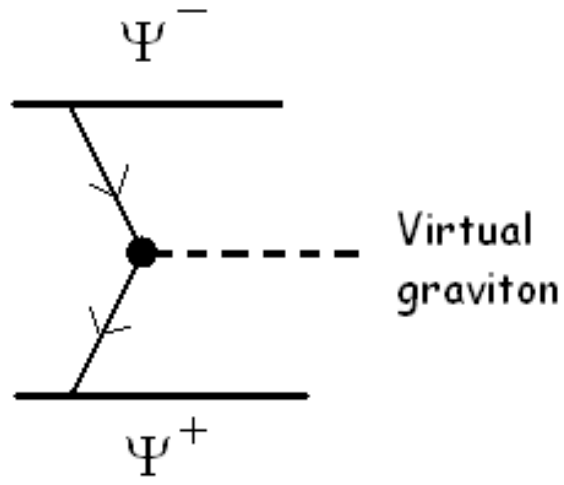

Fig. 4. Emission of a virtual graviton with spin 1 in the spontaneous decay $\psi^- \to \psi^+$. The matrix element of the mass-dipole moment operator between $\psi^-$ and $\psi^+$ has module $Mr/2$.

It is straightforward to check that there is an oscillating mass dipole between the states $\psi^-$ and $\psi^+$:

$$\begin{aligned}\langle \psi^+ | \hat{\mathbf{d}} | \psi^- \rangle &= \frac{1}{\sqrt{2}}\left(\langle 1| + \langle 2|\right) \hat{\mathbf{d}} \frac{1}{\sqrt{2}}\left(|1\rangle - |2\rangle\right) = \\ &= \frac{1}{2}\left(\langle 1| + \langle 2|\right)\left(M_1 \mathbf{r}_1 |1\rangle - M_2 \mathbf{r}_2 |2\rangle\right) = \frac{1}{2} M\mathbf{r}\end{aligned} \tag{23}$$

where $\mathbf{r}_1 = \frac{1}{2}\mathbf{r}$, $\mathbf{r}_2 = -\frac{1}{2}\mathbf{r}$; here $\mathbf{r}$ is the displacement between the masses $M_1$ and $M_2$, which in the end are taken to be equal. The origin of the coordinate system is in the center of mass.

This mass dipole moment has purely quantum origin, because in our system there are no masses of different signs, and it is known that in this case the classical mass dipole moment computed with respect to the center of mass is zero. We could say that the non-zero matrix element (23) is due to the quantum tunnelling between the states $|1\rangle$ and $|2\rangle$. This corresponds to a mass oscillation.

Eq. (22) gives the lifetime $\tau$ of the excited level $\psi^-$ by spontaneous emission. With the values of $M$ and $r$ found in Section 3.1 supposing an excitation frequency of the order of 1 MHz, one finds $B \approx 10^{12}$ m³/Js² for the stimulated emission coefficient and $\tau = A^{-1} \approx 1$ s for the lifetime for spontaneous emission (taking $\lambda f \approx 1$ m/s: compare discussion in (Modanese, 2011) and Sect. 5). The general dependence of $B$ on the frequency $\omega$ and on the length $r$ of the dipoles is easily obtained from eq.s (20), (22) and (23):

$$B \approx \frac{1}{\hbar}\omega r^3 \tag{24}$$

Note that $B$ is independent from the Newton constant $G$.

### 3.3 Digression: elementary dynamics of virtual particles with negative mass

Real particles with negative mass cannot exist, because they would make the world terribly unstable, popping up spontaneously from the vacuum with production of energy. In this work, however, we hypothesize the existence of long-lived virtual particles with negative mass, whose creation from the vacuum does not require or generate any energy. We recognize that these virtual particles have negative mass by looking at their metric at infinity, which is Schwarzschild-like, but with negative $M$ and negative ADM energy. We know that the dynamics of virtual particles, after their creation, is similar to that of real particles, and we have computed quantum amplitudes involving them.

We do not know any general principle about the "classical" dynamics of virtual particles with negative mass. Actually, virtual particles of this kind are an emergent phenomenon guessed from the path integral and can only be observed in a very indirect way. It is interesting, nonetheless, to make some reasonable hypothesis and check the consequences. Our basic assumption will be the following: for an isolated system comprising particles with positive and negative mass, the position of the center of mass, defined by

$$\mathbf{r}_{CM} = \sum_i M_i \mathbf{r}_i \qquad (25)$$

is invariant in time. From this assumption one can prove in a straightforward way several strange properties of particles with negative mass. These properties can be summarized by saying that in the usual dynamical rules their mass really behaves like a negative number, namely: (a) The acceleration of the virtual particle is opposite to the applied force. (b) The momentum is opposite to the velocity. (c) The kinetic energy is negative. The kinetic energy is defined as usual through the work of the applied force, in such a way that the sum $E_{\text{kin}}+E_{\text{pot}}$ is conserved.

Applying these rules one obtains a bizarre behaviour in the scattering processes and in the dynamics. For instance, although the gravitational potential energy of two virtual particles with negative mass is negative, $E_{\text{pot}} = -GM_1M_2/r$ (compare Sect. 3.1), the two particles experience a repulsion, due to Property (a). They tend to run away from each other; while their distance increases, their $E_{\text{pot}}$ decreases in absolute value, and their (negative) $E_{\text{kin}}$ increases in absolute value. If the particles were initially at rest at some distance $r_0$ (Fig. 5), when their distance goes to infinity they gain a $E_{\text{kin}}$ equal to their initial $E_{\text{pot}}$.

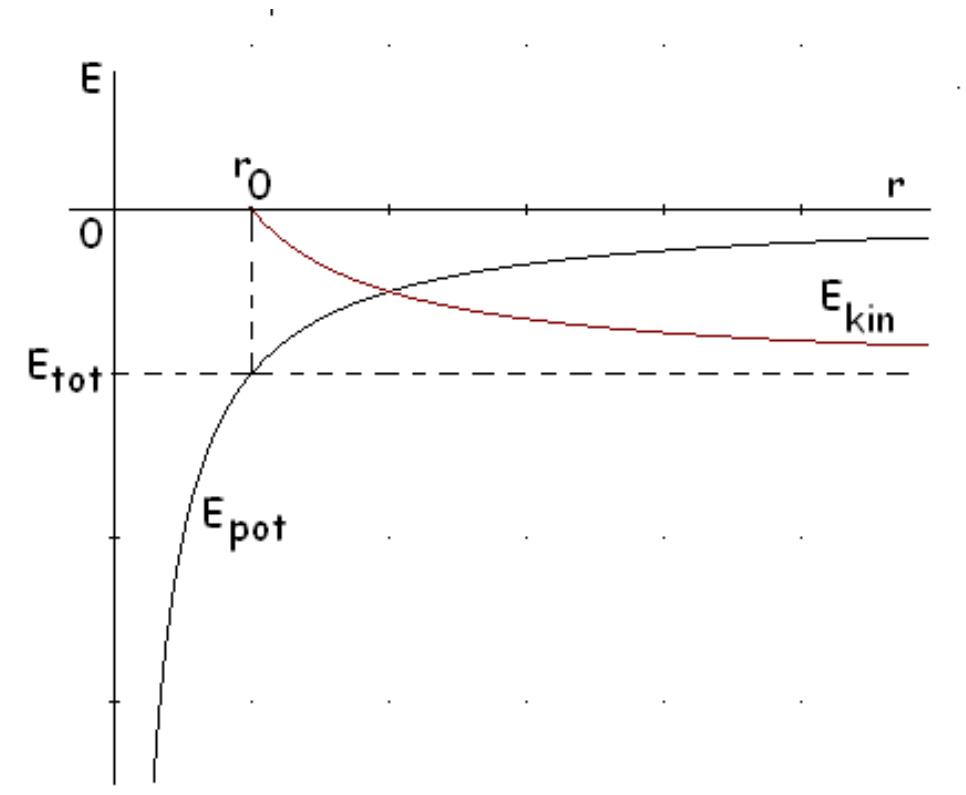

Fig. 5. "Classical" motion of two virtual particles with negative mass initially at rest at distance $r_0$. Although their potential energy is negative, they feel a repulsion and their (negative) kinetic energy increases in absolute value as their distance goes to infinity.

In the decay $\psi^- \rightarrow \psi^+$ (Sect. 3.2) the momentum of the emitted graviton is balanced by the recoil of the zero-modes (in the same direction of the emission). The conservation equations give

$$\begin{cases} Mv_r^2 + E_g = \Delta E \\ 2Mv_r - p_g = 0 \end{cases} \tag{26}$$

where $\Delta E$ is the energy gap, $E_g$ and $p_g$ are the graviton energy and momentum, $v_r$ is the recoil velocity of the zero-mode and $2M \approx -10^{-13}$ kg is the zero-mode mass. After replacing $p_g = \alpha E_g$, the system (26) leads to the equation

$$\frac{1}{M\alpha^2} E_g^2 + E_g - \Delta E = 0 \tag{27}$$

which has a positive solution $E_g \approx \Delta E$, independently from $\alpha$. Furthermore, the recoil velocity $v_r$ turns out to be always non-relativistic. This means that the recoil of the zero-modes can always ensure conservation of momentum, independently from the value of the graviton energy-momentum ratio $\alpha$.

## 4. Interaction of the zero-modes with a variable $\Lambda$-term

In Sect. 3 we have computed the probability of the decay process $\psi^- \rightarrow \psi^+$ with emission of a virtual graviton. The excitation process $\psi^+ \rightarrow \psi^-$ (transition of a zero-modes pair from a symmetric to an anti-symmetric state) can occur by absorption of a virtual graviton or by coupling to an external source. It is easy to show (Sect. 4.3) that the coupling of zero-modes to "ordinary" matter with energy-momentum $T_{\mu\nu} \propto m \frac{dx_\mu}{d\tau}\frac{dx_\nu}{d\tau}$ is exceedingly weak.

(Note that certain interactions between zero-modes and massive particles vanish exactly for symmetry reasons. For instance, a particle in uniform motion can never "lose energy in collisions with the zero-modes", because in its rest reference system the particle will see the vacuum, zero-modes included, as homogeneous and isotropic. There are possible exceptions to this argument: accelerated particles, or particles in states with large $p$ uncertainty.)

The coupling to a $\Lambda(t)$ term, or local time-dependent vacuum energy density, can lead to a significant transition probability. This is due to the presence of the non-linear $\sqrt{g}$ factor in the coupling, and corresponds physically to the fact that such a $\Lambda$ term does not describe isolated particles, but coherent, delocalized matter.

### 4.1 Summary of conventions and of some previous results

The Einstein equations with a cosmological constant, or vacuum energy term, are written

$$R_{\mu\nu} - \frac{1}{2} g_{\mu\nu} R + \Lambda g_{\mu\nu} = -\frac{8\pi G}{c^4} T_{\mu\nu} \tag{28}$$

The corresponding action (without the boundary term) is

$$S_E = -\frac{c^4}{8\pi G} \int d^4x \sqrt{g} R + \frac{\Lambda c^4}{8\pi G} \int d^4x \sqrt{g} \tag{29}$$

In this paper with use metric signature (+,-,-,-). With this convention, the cosmological (repulsive) background experimentally observed is of the order of $\Lambda c^4/G$=-$10^{-9}$ J/m$^3$.

In perturbative quantum gravity on a flat background, this value of $\Lambda$ corresponds to a small real graviton mass (Datta *et al.*, 2003, and ref.s). Actually, in the presence of a curved background the flat space quantization must be replaced by a suitable curved-space quantization (Novello & Neves, 2003). The limit $m \to 0$ of a theory with massive gravitons is tricky, so this global value of $\Lambda$ still represents a challenge for quantum gravity (besides the need to explain its origin; compare Sect. 2.1).

In our previous work we introduced the idea that at the *local* level, the coupling of gravity with certain coherent condensed-matter systems could give an effective local positive contribution to the cosmological constant and therefore generate instabilities in the field (imaginary graviton mass (Modanese, 1996)). This early argument was not very compelling, but was reinforced after considering the effects of the $\Lambda$-term on the weak-field dipolar fluctuations mentioned in Sect. 2.1. Still such effects were predicted to be very weak and dependent on the sign and value of the background $\Lambda$ at the scale of interest. After the discovery of the strong-field zero-modes of the action, in (Modanese & Junker, 2007) we computed the effect of a Λ-term on such configurations, but it still turned out to be very small.

#### 4.2 Time-dependent Λ and zero-modes transitions

A substantial progress was made in (Modanese, 2011), where we showed that the effect of a high-frequency Λ(t)-term can be quite large and independent from its sign and from the background Λ. This was obtained considering the *interactions* between the zero-modes, as we detail in the following. Our latest computations also allow us to recognize more clearly the difference between the gravitational effect of coherent matter mediated by the Λ-term and the (negligible) gravitational effect of the classical $T_{\mu\nu}$ of the same matter. After writing the total Lagrangian $L_{grav}+L_{matter}$, we split $L_{matter}$ into an "incoherent particles" part (Sect. 4.3) and a coherent matter part, described by a scalar field ϕ. Only the latter part contains a nonlinear factor $\sqrt{g}$, which can have non-vanishing matrix elements already to first order in Λ.

We suppose that the scalar field ϕ which describes the coherent matter has in flat space an action of the standard form

$$S_\phi = \int dx L_\phi = \int dx \left( \frac{1}{2}\partial^\mu \phi \partial_\mu \phi - \frac{1}{2} m_\phi^2 \phi^2 + k\phi^4 \right) \tag{30}$$

The gravitational coupling introduces a $\sqrt{g}$ volume factor. The dynamics of ϕ is driven by external forces, so this coupling can be regarded as an external perturbation $H_\Lambda$ , a local vacuum energy density term due to the presence of coherent matter described by a macroscopic wavefunction equivalent to a classical field:

$$H_\Lambda(t,\mathbf{x}) = \frac{1}{8\pi G}\sqrt{g(t,\mathbf{x})}\Lambda(t,\mathbf{x}) = \frac{1}{8\pi G}\sqrt{g(t,\mathbf{x})}L_\phi(t,\mathbf{x}) \tag{31}$$

The Λ term in coherent matter turns out to be much larger than the cosmological background: for instance, one typically has $\Lambda c^4/G=10^6\text{-}10^8$ J/m$^3$ in superconductors, depending on the type, while the currently accepted value for the cosmological background is of the order of $\Lambda c^4/G=10^{-9}$ J/m$^3$. The value above for superconductors is the result of a complex evaluation of the relativistic limit of the Ginzburg-Landau Lagrangian, which yields the following expression for Λ in terms of the pairs density ρ (Modanese, 2003):

$$\Lambda(t,\mathbf{x}) = -\frac{1}{2m}\left[\hbar^2(\nabla\rho)^2 + \hbar^2\rho\nabla^2\rho - m\beta\rho^4\right] \tag{32}$$

where β is the second Ginzburg-Landau coefficient and $m$ is the Cooper pair mass. This energy density has strong variations in space and time, following the behaviour of the macroscopic wavefunction. In order to obtain high-frequency oscillations in time, one can induce in the material Josephson currents (Modanese & Junker, 2007). Spatial variations have a typical scale of 1 nm, so we take this as the size of the volume $V$ where the perturbation is spatially constant and the transition probability is computed. For this reason we shall leave only the time dependence in $\Lambda$ and write henceforth

$$H_\Lambda(t,\mathbf{x}) = \frac{1}{8\pi G}\sqrt{g(t,\mathbf{x})}\Lambda(t) \tag{33}$$

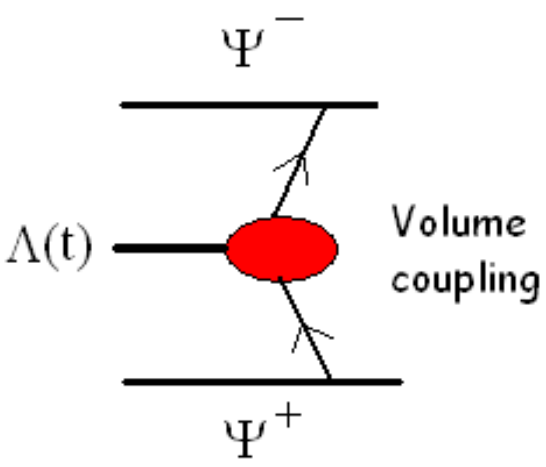

Fig. 6. A time-dependent $\Lambda$-term can be quite efficient in exciting transitions $\psi^+\to\psi^-$, because it enters the matrix elements to first order in $\Lambda$. The denomination "Volume coupling" refers to its mathematical form and to the fact that it is due to de-localized coherent matter described by a macroscopic wavefunction.

For the evaluation of the density of final states we refer to (Modanese, 2011) and give here only the final result on the probability of transitions $\psi^+\to\psi^-$ under the action of an external perturbation with oscillation frequency $\omega$ in resonance with the energy difference (20). Given the large number of available states, the resonance occurs for any frequency, and also if the perturbation is not exactly monochromatic.

In accordance with the Fermi rule and considering a volume $V\approx 10^{-27}$ m$^3$ and a frequency $\omega\approx 10^7$ Hz (compare Sect. 3), we obtain

$$\frac{dP}{dt} = \frac{1}{\hbar}\left|\left\langle \psi^+ \middle| H_\Lambda \middle| \psi^- \right\rangle\right|^2 \rho(E) \approx 10^{34}10^{-38}10^{27} \approx 10^{23}\mathrm{s}^{-1} \tag{34}$$

This implies that the excitation time of the zero-modes in the presence of a suitable local $\Lambda$-term is very short ($10^{-23}$ s). It is likely, actually, that this excitation process is limited by the energetic balance rather than by the transition probability.

### 4.3 Comparison with the effect of incoherent matter

The action of free incoherent particles is

$$S = \sum_a m_a \int \sqrt{g_{\mu\nu}(x_a)dx_a^\mu dx_a^\nu} \tag{35}$$

The index "$a$" denotes the sum over particles and will be omitted in the following, considering for simplicity one single particle. The corresponding field/particle interaction Hamiltonian density is

$$H_{\mathbf{x},particle} = \frac{1}{2m} h_{ij}(t,\mathbf{x}) p^i p^j \qquad (36)$$

where $m$ is the particle mass, $h_{ij} = g_{ij} - \eta_{ij}$ and $i,j$ are spatial indices. This holds to lowest order in $p$ and for fields $h$ which describe a plane wave (on-shell or off-shell, see proof in (Modanese, 2011)).

Suppose to apply eq. (36) to our case, i.e. to compute a transition probability $\psi^+ \rightarrow \psi^-$ due to the coupling of gravitation to single particles in ordinary matter. In this case, the particle momentum is a given numerical function of time, while $h_{ij}(t,\mathbf{x})$ is a quantum operator which acts on the Fock vacuum creating or destroying a graviton. (In the following we shall often denote the field operator as $\hat{h}$ and omit the indices.) The numerical factor $\frac{p^i p^j}{2m}$ is of the order of the kinetic energy of the particle $\frac{p^2}{2m}$, i.e. of the order of $10^{-19}$ J for an atomic system. This is also the magnitude order of the $\Lambda(t)$ term. But while the interaction Hamiltonian $H_{\mathbf{x},\Lambda}$ can have non-vanishing matrix elements also when acting linearly between the states $\psi^+$ and $\psi^-$, because it is proportional to the non-linear function $\sqrt{g} = 1 + \mathrm{Tr}(h) + \ldots$, the Hamiltonian $H_{\mathbf{x},particle}$ has non-vanishing matrix elements only to second order.

Namely, we can write a matrix element of the form $\langle In|\hat{h}|Out\rangle$ as

$$\langle In|\hat{h}|Out\rangle = \langle 0, In_{z-m}|\hat{h}|0, Out_{z-m}\rangle \qquad (37)$$

where $|In_{z-m}\rangle$ and $|Out_{z-m}\rangle$ denote the zero-mode components of the initial and final states, and $|0\rangle$ denotes the Fock vacuum, without gravitons. The matrix element is clearly zero, because it contains a single field acting between two Fock vacuum states. In other words, we can say that since neither in the initial state nor in the final state there are gravitons, the standard vertex (36) can have non-zero matrix element only when it is taken twice (Fig. 7) and is therefore proportional to $(p^2/m)^2$; but this is of magnitude order $10^{-38}$ in S.I. units, as seen, and gives a factor $10^{-76}$ after insertion in the transition probability (34).

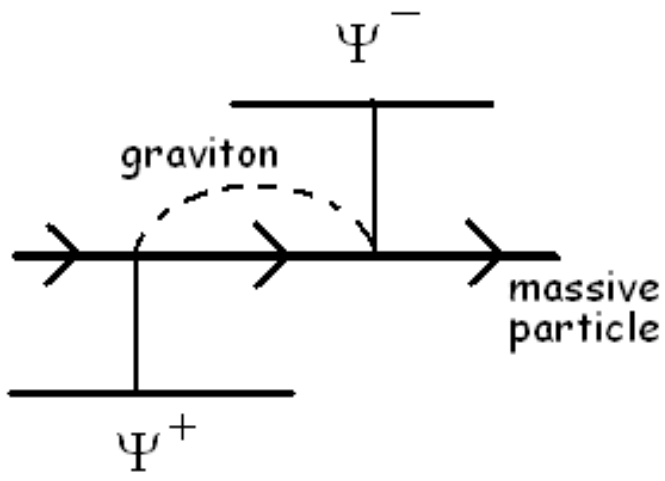

Fig. 7. Zero-mode excitation by double interaction with incoherent matter. A single massive particle can cause such an excitation by emitting and re-absorbing a virtual graviton, but the probability of this process is very small.

On the other hand, the $\hat{h}\hat{h}$, $\hat{h}\hat{h}\hat{h}$ , … terms in the expansion of $H_\Lambda$ can give non-zero matrix elements already to first order in Λ. We are not able to compute these matrix elements without a complete theory, because inside the Schwarzschild radius of the zero-modes the weak field expansion is not valid. The situation resembles that of early nuclear physics, where the nuclear matrix elements were largely unknown, apart from some general properties or magnitude orders; this did not prevent researchers from obtaining important data on the processes, based on the available information and on the crucial knowledge of the final states density.

## 5 Properties of virtual gravitons

The aim of this final section is to give a simplified yet consistent physical picture of how virtual gravitons mediate the gravitational interactions. This is necessary in order to understand the link between virtual gravitons and the other kind of vacuum fluctuations studied in this paper, the zero-modes.

Note that virtual gravitons respect the usual time-energy uncertainty principle; their are not "long-lived" vacuum fluctuations like the zero-modes. This is because we consider gravitons as the particles obtained in the perturbative quantization of gravity on a flat background. It is known that the theory is not renormalizable at higher orders, but we use only tree diagrams in this work and suppose that the renormalization problem will be solved or is already solved in an effective quantum field theory of gravity (compare Sect. 1).

The concept of virtual particles mediating an interaction is not simple, and it is sometimes used improperly. In some treatments the virtual particles are seen as purely formal representations of perturbative diagrams. Instead, it is important to understand in which sense they can be regarded as particles or not.

For a real particle of given mass $m$, kinematics allows to connect the three quantities $E$, $p$, $v$ through the two relations

$$E^2 - p^2c^2 = m^2c^4 \quad (38)$$

$$E = \frac{mc^2}{\sqrt{1 - v^2 / c^2}} \quad (39)$$

Therefore when one of the tree quantities is known, we can find the other two. Note that from (38) and (39) one can prove the relation $p / E = v / c^2$, which connects $E$ and $p$ and (unlike (39)) also holds for $v=c$. So we can as well consider as basic relations between $E$, $p$, $v$ the couple

$$E^2 - p^2c^2 = m^2c^4 \quad (40)$$

$$\frac{p}{E} = \frac{v}{c^2} \quad (41)$$

These formulas all hold when the quantities $m$, $E$, $p$, $v$ are well defined, thus for particles which are either stable or have a sufficiently long lifetime. For virtual particles the situation is more vague and one finds a range of statements in the literature. For instance, there is a simple textbook argument showing that the exchange of virtual photons gives rise to a $1/r^2$ force between two charges $q_1$ and $q_2$. The argument is based on the time-energy uncertainty relation. One writes $\Delta E \Delta t \approx h$, where $\Delta E$ is the energy of the exchanged virtual photon and $\Delta t$ its lifetime. Supposing that the virtual particle travels with light speed, its range is $r=c\Delta t$. Therefore if the charges $q_1$ and $q_2$ are at a distance $r$, the "exchanged energy" is $\Delta E \propto 1/r$ and the corresponding force will be proportional to $1/r^2$. One must add the assumption that the number of exchanged photons is also proportional to the product $q_1 q_2$ of the charges of the interacting particles. A weak point in this argument is the identification of the exchanged energy with the potential energy of the interaction. In fact, the exchanged energy depends on the velocities of the charged particles and can even be zero for static sources or in cases like that of the protons observed in their center of mass system (Fig. 8, Sect. 5.1). Apart from this, the assumption that the virtual particle has an energy uncertainty and that it propagates with light speed looks reasonable.

### 5.1 Example: scattering process

Let us consider, however, another simple example: the electromagnetic scattering of two protons with the exchange of a single virtual photon. To fix the ideas, we choose a definite energy of the two protons as seen in their center of mass system, for instance $E=10^{-13}$ J ≈ 1 MeV. (Magnitude orders are important in these considerations, in order to estimate the wavelength and the number of the exchanged particles, as we shall see below in the case of gravitons.) In this reference system the exchanged energy is zero and the exchanged momentum is of the order of $\sqrt{2m_pE} \approx 10^{-20}$ kg m/s (non-relativistic approximation).

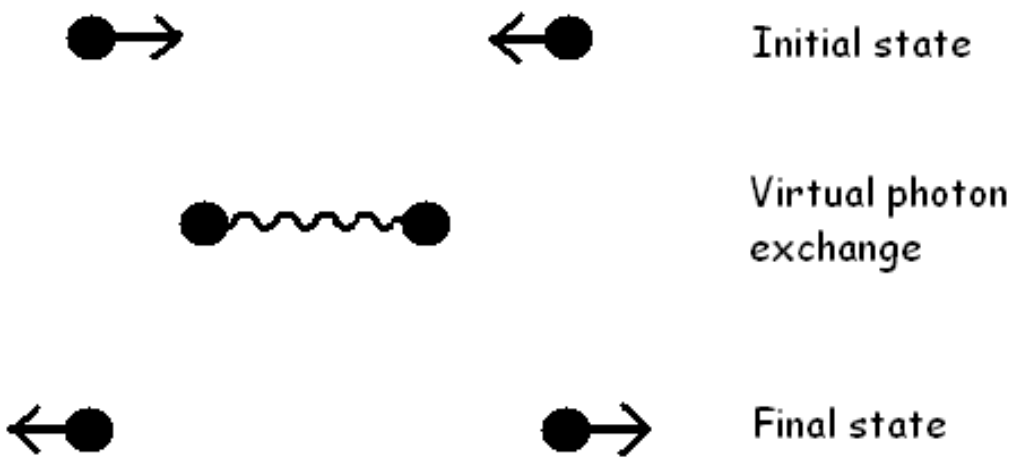

Fig. 8. Proton-proton scattering through the exchange of a single virtual photon, as seen from the center of mass reference system.

Suppose that this momentum is carried by one single virtual photon γ. The photon is off-shell, with imaginary mass $m_\gamma^2<0$: $m_\gamma^2 c^2 = E_\gamma^2 - p_\gamma^2 c^2 \Rightarrow m_\gamma^2 = -p_\gamma^2/c^2$. The virtual photon energy and momentum are exactly defined and their ratio $E_\gamma/p_\gamma$ is exactly zero in this reference system (it is not Lorentz-invariant). The wavelength of the photon, defined as $\lambda=h/p_\gamma$, is of the order of $10^{-14}$ m. One can estimate, classically, that the minimum distance reached by the protons is of the order of $10^{-16}$ m. If the virtual photon is emitted at this point, its wavefunction can clearly not be regarded as a plane wave. Its propagation velocity $v$ is hardly observable and relation (41) appears to suggest that $v$ is very large; if we assume $v=c$, it is only by analogy with the familiar retarded classical effects.

The situation appears, in conclusion, to be very different from the previous example. It seems reasonable to draw a clear distinction between a scattering process, which can be described as the exchange of a single virtual particle, and the inter-particle force in static or quasi-static conditions, which is in general equivalent to the exchange of a large number of virtual particles.

### 5.2 Photons or gravitons vs. static force

Let us now consider a different situation (Fig. 10): a massive particle (for instance, a proton) in free fall in the gravitational field of the Earth. Suppose the particle is initially at rest. There is an exact quantum formula which allows to find the static interaction potential energy in field theories like QED, QCD etc. The generalization to quantum gravity was given by (Modanese, 1995). In this formula the graviton propagator appears explicitly, as well as the $G$ constant and the masses $m_1 m_2$ of the sources (showing that the amplitude of virtual gravitons generation is proportional to both these masses; this property was also discussed by (Clark, 2001)). The potential energy is written as

$$U(r) = m_1 m_2 \lim_{T\to\infty} \frac{1}{T} \int_{-T/2}^{T/2} dt_1 \int_{-T/2}^{T/2} dt_2 \left\langle 0 \left| h_{00}(t_1, \mathbf{r}_1) h_{00}(t_2, \mathbf{r}_2) \right| 0 \right\rangle \qquad (42)$$

This equation describes the exchange of gravitons, for an ideally infinite time, between two static masses ($\mathbf{r} = \mathbf{r}_1 - \mathbf{r}_2$; see Fig. 9). In our case the masses are the Earth and the particle.

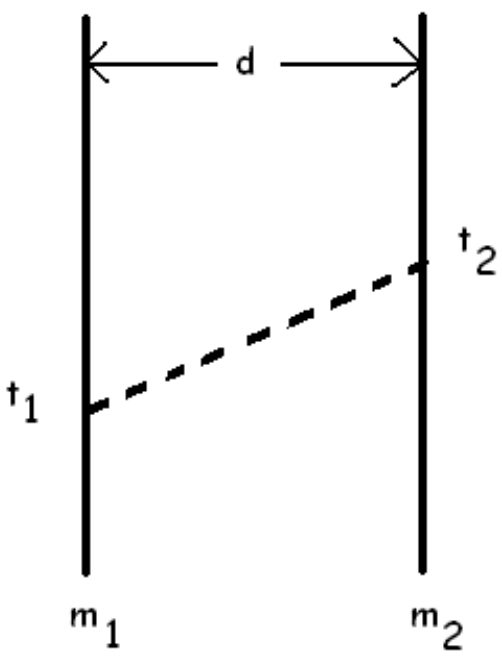

Fig. 9. Static potential energy of two masses $m_1$, $m_2$ as the outcome of graviton exchange. Virtual gravitons are emitted and absorbed at all possible times $t_1$, $t_2$; the final result is obtained by integration over $t_1$ and $t_2$.

The gravitons flux is proportional to both $m_1$ and $m_2$ and the propagator gives the amplitude of the propagation of virtual gravitons from $\mathbf{r}_1$ to $\mathbf{r}_2$, but note that their emission and absorption probabilities are equal to 1. If we expand the Feynman propagator in four-momentum space, we can see which energies and momenta are exchanged. One first finds

$$U(r) = Gm_1 m_2 \lim_{T\to\infty} \frac{1}{T} \int_{-T/2}^{T/2} dt_1 \int_{-T/2}^{T/2} dt_2 \int dE \int d^3p \frac{e^{i\mathbf{pr} - iE(t_1 - t_2)}}{E^2 - \vec{p}^2 - i\varepsilon} \qquad (43)$$

Changing variables to ($t_1$+$t_2$), ($t_1$-$t_2$) we find that the integral in ($t_1$+$t_2$) cancels the factor $1/T$. By integrating exp[-$iE(t_1-t_2)$] one obtains δ($E$): this selects the static limit, i.e. the exchanged gravitons have $E \cong 0$ (note that in eq.s (43) and (44) we use natural units $h/2\pi = c = 1$). Finally we have

$$U(r) = -Gm_1 m_2 \int d^3p \frac{e^{i\mathbf{pr}}}{p^2} = -\frac{2Gm_1 m_2}{\pi r} \int_0^\infty dp' \frac{\sin p'}{p'} \qquad (44)$$

with $p' = |\mathbf{p}| r$. (A similar reasoning also applies to quantum electrodynamics.) The last integral is equal to $\pi/2$ and the main contribution to the integration comes from the momentum region $p' < \pi$, i.e. $pr < \pi$. This means that in the classical interaction of two masses at distance $r$, the majority of the exchanged virtual gravitons have momentum $p < \hbar / r$ (restoring $\hbar$), or wavelength $\lambda > r$.

The propagation velocity is not the same for all virtual gravitons, as is seen from the fact that their emission/absorption times vary from -∞ to +∞; correspondingly, their invariant masses also vary. Being a static formula, eq. (42) cannot show that the propagation velocity of the force is *c*. For this we need some generalization to moving sources; the formalism of Quantum Field Theory will ensure that the retardation effects are accounted for.

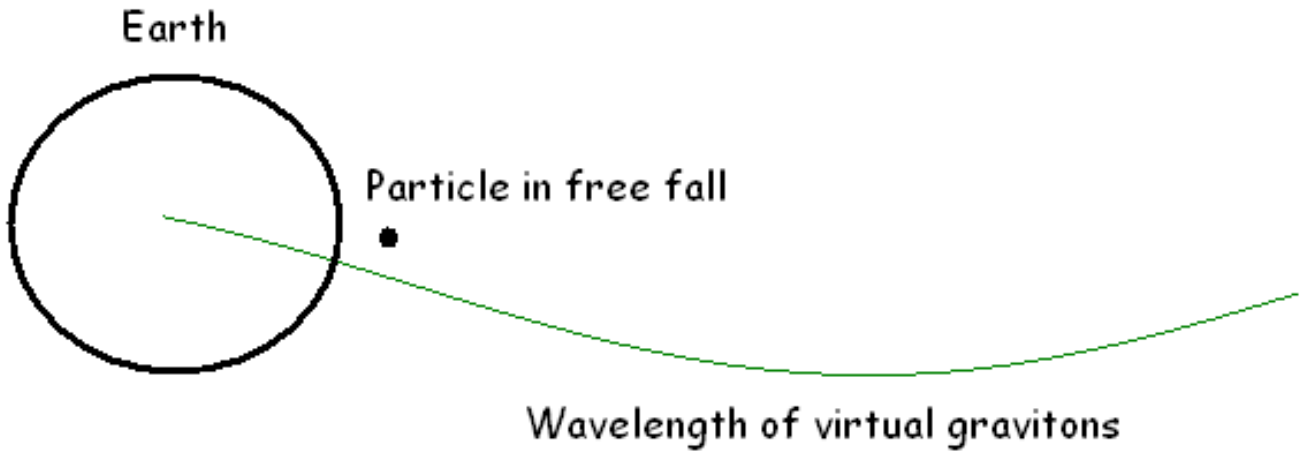

Fig. 10. The wavefunction of virtual gravitons exchanged in a quasi-static interaction is not a plane wave: for most of them, the wavelength is larger than the traveled distance.

The condition $p < \hbar / r$ or $\lambda > r$ shows that the wavefunctions of the exchanged virtual gravitons are very different from plane waves: these functions do not even make a complete oscillation over a distance equal to the Earth radius! Such virtual gravitons can hardly be regarded as "particles". This should actually be expected, because the attractive character of the force can only be understood if we consider the details of the wavefunction (Baez, 1995). A naïve particle exchange can only explain repulsive forces.

Let us estimate how many gravitons are exchanges between the Earth and a free-falling proton or nucleon. To stay close to the static limit, consider a short interval $\Delta t$. The proton, initially at rest, acquires in this time a momentum $p=mg\Delta t$. The absorbed gravitons have, on the average, individual momentum $p \approx \hbar / r \approx 10^{-41}$ kg m/s. The number of absorbed gravitons is then of the order of $N/\Delta t \approx 10^{15}$ $s^{-1}$.

### 5.3 Virtual gravitons emitted in the decay $\psi^- \to \psi^+$

We have seen that the virtual gravitons exchanged in a quasi-static attractive gravitational interaction have very small energy and momentum. Their wavefunctions do not resemble plane waves. The propagation of this "stream" composed of a large number of virtual gravitons is a collective phenomenon occurring at light velocity.

The virtual gravitons emitted in the decay $\psi^- \to \psi^+$ (Sect. 3.2) have completely different features. Their energy is much larger (≈$10^{-27}$ J). Their momentum is not fixed by the emission process, since the recoil of the emitting zero-modes can balance it in any case. One of these gravitons can be individually absorbed by a real particle at rest (for instance a proton), in such a way to conserve energy and momentum, provided the product $f\lambda$ of the graviton frequency and wavelength is equal to half the final velocity of the real particle. In fact the balance equations are

$$\frac{hf}{\frac{h}{\lambda}}=\frac{E_g}{p_g}=\frac{E_p}{p_p}=\frac{\frac{1}{2}mv_p^2}{mv_p}=\frac{1}{2}v_p \tag{45}$$

where the suffix "$g$" denotes the virtual graviton and "$p$" the real particle. This is a quantum process that satisfies the conservation balance, thus it can happen and will in fact happen, with a certain amplitude. The amplitude for the final step (absorption by the real particle) is unitary, by analogy with the similar process in the static exchange.

Supposing that the real particle is a proton, it is easy to check that conservation requires $v_p \approx$ 1 m/s, $\lambda \approx 10^{-7}$ m. If the distance between the real particle and the graviton source is much larger than $\lambda$, then the wavefunction of the virtual graviton can be properly described as a plane wave. If it is legitimate to apply the kinematical relations of Sect. 5.1 to this plane wave, it follows that the virtual graviton propagates like a tachyon. This does not violate the causal principles of special relativity, because the propagation of a single virtual particle cannot be modulated to obtain a signal. The existence of such tachyonic virtual gravitons would be a consequence of the unique features of their source (virtual decay $\psi^- \to \psi^+$).

## 6. Conclusions

In quantum gravity the vacuum fluctuations have a more complex structure than in other field theories with positive-definite action. In particular, there are vacuum fluctuations which in the non-interacting approximation have infinite lifetime, and seen from the outside appear as Schwarzschild metrics with negative mass. These vacuum fluctuations behave as pseudo-particles which are created "for free" from the vacuum at any point in spacetime. The non-interacting vacuum can in fact be described as an incoherent, homogeneous and isotropic superposition of a Fock vacuum plus infinite states of this kind ("zero-modes").

When the interaction is taken into account, one finds that each pair of zero-modes with equal virtual mass $M$ and distance $r$ can be in two states, denoted by $\psi^+$ and $\psi^-$, with energy splitting $\Delta E=E^- - E^+=GM^2/r$. The excited state $\psi^-$ can decay into the state $\psi^+$ by emitting a virtual off-shell graviton with spin 1. The energy-momentum ratio $E/p$ of the virtual graviton can take in principle any value, being the total momentum preserved by the recoil of the zero-modes pair. The $A$ and $B$ Einstein coefficients of spontaneous and stimulated emission have been computed in weak-field approximation. The $B$ coefficient turns out to be of the order of $\omega r^2/2\pi h$, where $\omega$ is the frequency corresponding to the gap $\Delta E$. The $A$ coefficient depends on the wavelength; for $\omega\lambda \approx 1$ m/s one has $A \approx 1$ s$^{-1}$.

The excitation process $\psi^+ \to \psi^-$ cannot occur by interaction with single incoherent particles, because the relative amplitude is exceedingly small, involving a double elementary particle/graviton vertex. Instead, a sizeable excitation amplitude is obtained in the interaction with an external source of the form $\int dx\sqrt{g}\Lambda(t)$ (local vacuum energy density term, due to the presence of condensed matter in a coherent state). By taking into account the density of final states one finds, for a length scale of the Λ-term of the order of $10^{-9}$ m, an excitation time $\psi^+ \rightarrow \psi^-$ of the order of $10^{-23}$ s.

The virtual gravitons emitted in the decay $\psi^- \rightarrow \psi^+$ are very different from those exchanged in the usual gravitational interactions. Consider, for instance, a nucleon in free fall near the surface of the Earth. If it was initially at rest, it reaches a velocity of 1 m/s in approximately 0.1 s, absorbing $\approx 10^{14}$ virtual gravitons of very low frequency and large wavelength. For comparison, a single virtual graviton of frequency $10^7$ Hz emitted in a vacuum decay $\psi^- \rightarrow \psi^+$ can transfer the same momentum to the nucleon in a single quick absorption process.